\documentclass{article}
\usepackage{amsmath}
\usepackage{cite}
\usepackage[linesnumbered,ruled]{algorithm2e}
\DontPrintSemicolon 
\usepackage{graphicx}
\usepackage{amsmath}
\usepackage{lipsum}
\usepackage{multicol}
\usepackage{tabularx}
\usepackage{hyperref}
\hypersetup{
	colorlinks,
	citecolor=black,
	filecolor=black,
	linkcolor=black,
	urlcolor=black
} 

\makeatletter
\newcommand{\removelatexerror}{\let\@latex@error\@gobble}

\newcommand{\nosemic}{\renewcommand{\@endalgocfline}{\relax}}
\newcommand{\dosemic}{\renewcommand{\@endalgocfline}{\algocf@endline}}
\let\oldnl\nl
\newcommand{\nonl}{\renewcommand{\nl}{\let\nl\oldnl}}

\makeatother

\begin{document}
\title{Efficient Generation of Parallel Spin-images Using Dynamic Loop Scheduling}


\author{Ahmed Eleliemy, Ali Mohammed, and Florina M. Ciorba\\
	Department of Mathematics and Computer Science\\
	University of Basel, Switzerland}

\date{}
\maketitle
\clearpage
\tableofcontents
\clearpage


\sloppy
\begin{abstract}
\label{sec:abstract}
High performance computing~\mbox{(HPC)} systems underwent a significant increase in their processing capabilities. 
Modern HPC systems combine large numbers of homogeneous and heterogeneous computing resources. 
Scalability is, therefore, an essential aspect of scientific applications to efficiently exploit the massive parallelism of modern HPC systems. 
This work introduces an efficient version of the parallel spin-image algorithm~\mbox{(PSIA)}, called~\mbox{EPSIA}.
The \mbox{PSIA} is a parallel version of the spin-image algorithm~\mbox{(SIA)}. 
The \mbox{(P)SIA} is used in various domains, such as 3D object recognition, categorization, and \mbox{3D}~face recognition. 
\mbox{EPSIA} refers to the extended version of the~\mbox{PSIA} that integrates various well-known dynamic loop scheduling~\mbox{(DLS)} techniques. 
The present work: 
(1)~Proposes~\mbox{EPSIA}, a novel flexible version of \mbox{PSIA}; 
(2)~Showcases the benefits of applying~\mbox{DLS} techniques for optimizing the performance of the~\mbox{PSIA};
(3)~Assesses the performance of the proposed \mbox{EPSIA} by conducting several scalability experiments.
The performance results are promising and show that using well-known~\mbox{DLS} techniques, the performance of the~\mbox{EPSIA} outperforms the performance of the~\mbox{PSIA} by a factor of 1.2 and 2 for homogeneous and heterogeneous computing resources, respectively.
\end{abstract}


\newpage
\section{Introduction}
\label{sec:intro}

Modern high performance computing~\mbox{(HPC)} systems are characterized by a large number of computing resources and their heterogeneity.
Efficiently exploiting HPC systems along these two characteristics represents a significant challenge for parallel applications.
Increasing the number of computing resources assigned to a parallel application can reduce the execution time.
However, such a reduction is not guaranteed due to the management of parallelism and communication overheads. 
Also, heterogeneity adds several challenges in managing the assigned computing resources.
Executing parallel applications on heterogeneous resources requires that the effects of the lower performance computing resources do not dominate the performance of the other computing resources.

The parallel spin-image algorithm~\mbox{(PSIA)}~\cite{eleliemy2016loadbalancing} is a parallel version of the well known \mbox{spin-image} algorithm~\mbox{(SIA)}~\cite{CVJohnson}.
The \mbox{SIA} is widely used in different domains, such as face detection~\cite{choi2013angular}, object recognition~\cite{johnson1999using}, 3D map registration~\cite{mei2013new}, and 3D database retrieval systems~\cite{assfalg20043d}.
The main limitation of the \mbox{SIA} is its computational time complexity and, consequently, its execution time. 
The \mbox{PSIA}~\cite{eleliemy2016loadbalancing} is introduced to overcome this limitation of the \mbox{SIA}.
However, the \mbox{PSIA}~\cite{eleliemy2016loadbalancing} only employs a static load balancing technique to distribute the process of the \mbox{spin-image} generation among the available computing resources. 
Moreover, to the best of our knowledge, the scalability of the \mbox{PSIA} has not yet been studied.

Similar to most scientific applications, the main source of parallelism in the \mbox{PSIA} is a loop, which for \mbox{PSIA} consists of independent iterations.
Efficient loop scheduling techniques are, therefore, needed for parallelizing and executing this loop on parallel computing systems. 
Among the loop scheduling techniques, the dynamic loop scheduling~\mbox{(DLS)} techniques have been shown to be most effective in optimizing the execution of parallel loop iterations by scheduling them during execution~\cite{taxonomy,Plata,IBto,IBtog}.
This work proposes a novel version of \mbox{PSIA}, namely \mbox{EPSIA}, that integrates a number of \mbox{DLS} techniques, such as self scheduling \mbox{SS}~\cite{tang1986processor}, guided self scheduling \mbox{GSS}~\cite{polychronopoulos1987guided}, and factoring \mbox{FAC}~\cite{hummel1992factoring}.
 \mbox{EPSIA} is generic enough to integrate different \mbox{DLS} techniques.
The performance of the \mbox{PSIA} and the performance of the \mbox{EPSIA} are evaluated via different scalability experiments and on different homogeneous and heterogeneous computing resources. 
The achieved performance of the \mbox{EPSIA} under different \mbox{DLS} techniques is studied. 

The remainder of this work is organized as follows.
In Section~\ref{sec:background}, a background of the \mbox{PSIA} and the \mbox{DLS} techniques used in this work is provided. 
The most relevant work in the literature, concerning the performance optimizations of the \mbox{SIA}, is also reviewed in Section~\ref{sec:background}.
In Section~\ref{sec:proposed}, the proposed \mbox{EPSIA} that can use different \mbox{DLS} techniques is introduced. 
The experimental setup and the information needed to reproduce this work is presented in Section~\ref{sec:experimentsetup}. 
In Section~\ref{sec:results}, the results of executing the proposed \mbox{EPSIA} are compared with the results of executing the \mbox{PSIA} on homogeneous and heterogeneous computing resources to derive their weak and strong scalability profiles, respectively.
In Section~\ref{sec:conclusion}, the conclusion of this work and the potential future work are outlined.

\section{Background and Related Work}
\label{sec:background}
This section describes the PSIA and the most relevant work concerning the PSIA optimization.
\subsection{Parallel \mbox{Spin-Image} Algorithm}
\label{subsec:psia}

\begin{figure}
	\includegraphics[width=\columnwidth, clip,trim=0.3cm 15cm 0cm 1cm]{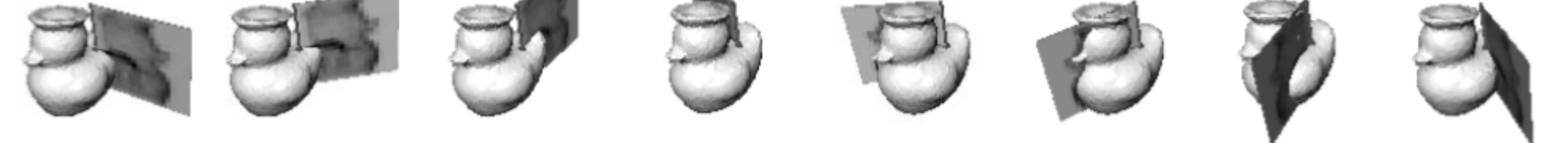}
	\caption{An analogy of the \mbox{spin-image} generation process with eight animation frames (from~\cite{CVJohnson})}
	\label{fig:spin}
\end{figure}

The spin-image algorithm~\mbox{(SIA)} was originally introduced in 1997 by Johnson~\cite{CVJohnson}.
It converts a \mbox{3D} object to a set of \mbox{2D} images which are considered as a shape descriptor for that \mbox{3D} object. 
The crux of the SIA is the process of generating the \mbox{2D} images. 
The \mbox{spin-image} generation process can be explained as a process of spinning a sheet of paper through a \mbox{3D} object. 
When a sheet of paper spins around a certain oriented point through a \mbox{3D} object, other oriented points of that object are pasted onto that sheet of paper. 
After a complete cycle around the oriented point, the spinning sheet of paper represents a \mbox{spin-image} generated at that oriented point.
In \figurename{~\ref{fig:spin}}, taken from~\cite{CVJohnson}, the \mbox{spin-image} generation process is illustrated using eight animated frames.

\begin{figure}
	\begin{center}
		\includegraphics[width=\textwidth, clip, trim= 0cm 21.5cm 5cm 0cm]{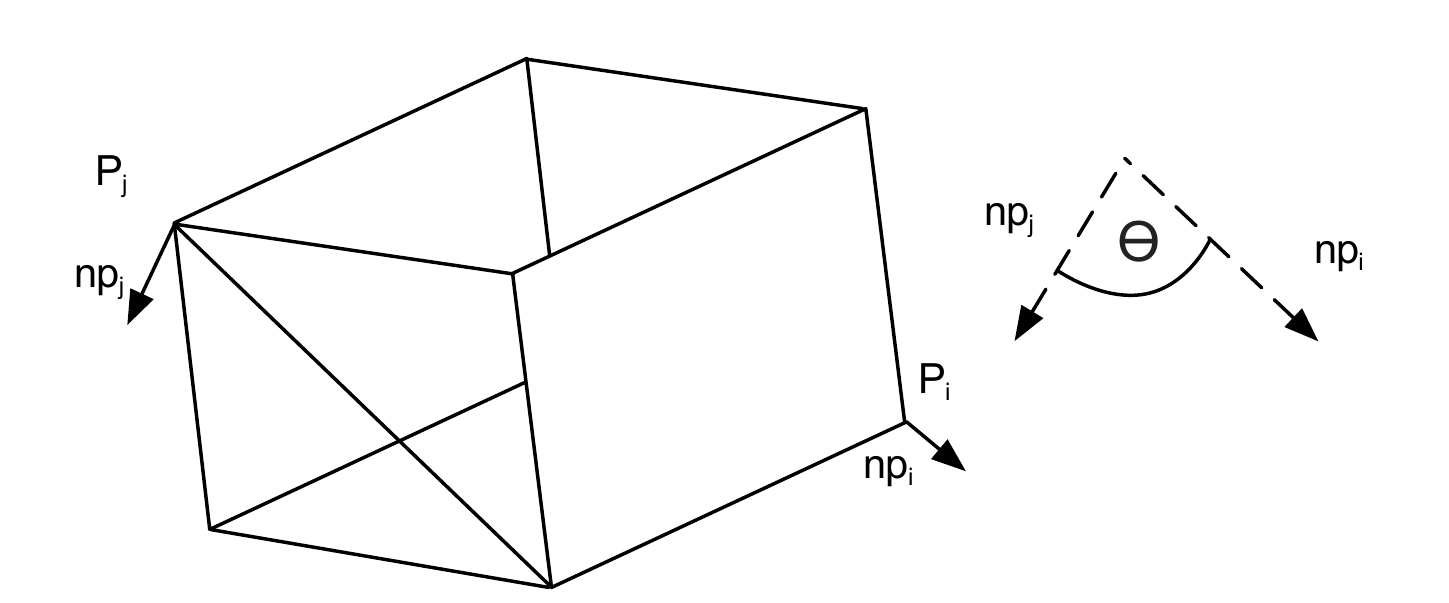}
		\caption{A \mbox{3D} object in which $P_j$ does not contribute to the generated spin-image at $P_i$ due to the fact that $\theta$ is greater than $S$}
		\label{fig:angle}
	\end{center}
\end{figure}

Three parameters characterize the~\mbox{SIA}:~$W$,~$B$, and~$S$, and are included in Table~\ref{tab:notation}.
$W$~denotes the number of pixels in a row or column of the generated \mbox{spin-image} and is similar to the width of the spinning sheet of paper.
The \mbox{SIA} assumes square \mbox{spin-images} with equal widths and heights.
$B$~is a factor of the 3D mesh resolution, used to determine the storage capacitance of each cell on the spinning sheet of paper.
Increasing~$B$ means that many oriented points will be pasted to the same cell on the spinning sheet of paper.
Consequently, the effect of individual oriented points on the generated \mbox{spin-image} will be reduced.
$S$~is a constraint for the \mbox{spin-image} generation process.
If the angle between $np_i$ and $np_j$, the normal vectors of the two oriented points $P_i$ and $P_j$, respectively, is greater than $S$, then the oriented point $P_j$ does not contribute to the generated \mbox{spin-image} at $P_i$. 
In \figurename{~\ref{fig:angle}}, $\theta$ is the angle between the two normal vectors $np_i$ and $np_j$.

\begin{table}
	\centering
	\vspace{-0.35cm}
	\caption{Glossary of Notation}
	\label{tab:notation}
	\begin{tabular*}{.9\columnwidth}{l|l}
	\textbf{Symbol} & \textbf{Description} \\
	\hline
	$M$   & \begin{tabular}[c]{@{}l@{}}  Number of oriented points \end{tabular} \\ \hline
	
	$N$  & \begin{tabular}[c]{@{}l@{}}Number of \mbox{spin-images}\\ $1 \leq N \leq M$ \end{tabular}\\ \hline
	
	$P_i$ & \begin{tabular}[c]{@{}l@{}} An oriented point with a known normal vector where \\ a \mbox{spin-image} can be generated, $0$ $\leq$ $i$  $\textless$ $M$ \end{tabular} \\ \hline
		
	$OP$ &  \begin{tabular}[c]{@{}l@{}}Set of all oriented points that belong to a \mbox{3D} object\\ \{$P_{i}$ $|$  $0$ $\leq$ $i$  $\textless$ $M$\} \end{tabular}     \\ \hline
		
	$np_i$ & A 3D vector that represents the normal vector of $P_i$\\ \hline
	
	$\theta$ & \begin{tabular}[c]{@{}l@{}} The angle between two normal vectors $np_i$ and $np_j$,\\ $0$ $\leq$ $i$,$j$  $\textless$ $M$ \end{tabular}\\ \hline
	
	$W$ & \begin{tabular}[c]{@{}l@{}} The number of pixels in a row or column of the \\generated \mbox{spin-image} where the generated \mbox{spin-image}\\ is assumed to be a square matrix \end{tabular}\\ \hline
		
	$B$   & \begin{tabular}[c]{@{}l@{}} A factor of the 3D mesh resolution that is used to \\determine the storage capacitance of the\\ generated \mbox{spin-image}, 0 $\textless$ $B$ $\leq$ 10 \end{tabular}\\ \hline
		
	$S$   & \begin{tabular}[c]{@{}l@{}}Maximum allowed angle $\theta$ between $P_i$ and $P_j$, where \\ $P_j$ contributes to the generated \mbox{spin-image} at $P_i$ \end{tabular}\\ \hline
		
	$WO$ &\begin{tabular}[c]{@{}l@{}} The number of workers used to generate the \mbox{spin-images}  \end{tabular}\\ \hline

	$wo_k$ &\begin{tabular}[c]{@{}l@{}} A worker that represents an MPI rank, pinned to a certain \\computing resource (core), $1 \leq k \leq WO$ \end{tabular}\\ \hline
	\end{tabular*}
\end{table}

The time complexity of the \mbox{SIA} is~$O(N M)$. 
If~$N$ approximately equals~$M$, \mbox{3D} objects with more than~$100K$ oriented points represent a significant challenge for the \mbox{SIA} in terms of its execution time.
PSIA~\cite{eleliemy2016loadbalancing} exploits the inherent parallelism within the \mbox{SIA} where the calculation of each individual \mbox{spin-image} is independent of other \mbox{spin-image} calculations. 
The steps of the \mbox{PSIA} are listed in Algorithm~\ref{algo:spin}.

\begin{figure}
	\removelatexerror
	\begin{algorithm}[H]
	 \SetKwInOut{Input}{Input parameters}
	 \SetKwInOut{Output}{Output parameter}
	 calculateSpinImages (W, B, S, OP, N)\;
	 \Input{\mbox{W: image width}, \mbox{B: bin size}, \mbox{S: support angle}, \mbox{OP: list of oriented points}, \mbox{N: number of generated spin-images}}
	 \Output{spinImages: list of generated spin-images}
	 spinImages = createSpinImagesList(N)\;
	 M = getLength(OP)\;
	 \textbf{Parallel} \For{  i = 0 $\rightarrow$ $N$}
	 { 
	 	tempSpinImage[W, W]\;
	 	init(tempSpinImage)\;
	 	P = OP[i]\;
	 	\For{j = 0 $\rightarrow$ $M$}
	 	{
	 		X = OP[j]\;
	 		$np_i$ = getNormal(P)\;
	 		$np_j$ = getNormal(X)\;
	 		\If{acos($np_i \cdot np_j$) $\le S$}
	 		{
	 	
	 			$k$  =  $\Bigg \lceil$ $\cfrac{W/2 - np_i \cdot (X-P) }{B}$ $\Bigg \rceil$ \;
	 			\vspace{0.2cm}
	 			$l$ =  $\Bigg \lceil$  $\cfrac{ \sqrt{||X-P||^2 - (np_i\cdot(X-P))^2} }{B}$  $\Bigg \rceil$\;
	 		     
	 		    \If{0 $\le$ k $\textless$ W and 0 $\le$ l $\textless$ W}
	 		    { tempSpinImage[k, l]++\;	}
	 		}
	 	}
	 	add(spinImages, tempSpinImage)\;
	 }
	 \caption{Parallel \mbox{spin-image} algorithm (from~\cite{eleliemy2016loadbalancing}) }
	 	\label{algo:spin}
\end{algorithm}
\end{figure}

There are two experimental setups for executing any implementation of Algorithm~\ref{algo:spin}. 
The first setup is when the number of parallel computing resources, $WO$, used in the experiment equals~$N$. 
In such a setup, each worker generates precisely one \mbox{spin-image}, i.e, according to Algorithm~\ref{algo:spin}, it executes the code between \mbox{Lines 5-20} \emph{only once}. 
In practice, it is not always feasible for the number of workers~$WO$ to equal~$N$, especially when~$N$ approximately equals $M$. 
The second setup is when $WO$ is smaller than~$N$.
Each worker generates a certain number of \mbox{spin-images} proportional to the ratio of $N$ divided by $WO$.
This means that each worker executes the code between \mbox{Lines 5-20} of Algorithm~\ref{algo:spin} \emph{more than once}.
In both experimental setups, the performance of the algorithm is dominated by the performance of the slowest worker.
A worker can be the slowest performing worker in two cases: (1)~It has a larger amount of computations than others and/or (2)~It has lower processing capabilities than others.

In \figurename{~\ref{fig:loadim}, twenty MPI ranks executed the PSIA application to generate  80,000 \mbox{spin-images}. They yielded unequal finishing times due to the workload imbalance.
According to Algorithm~\ref{algo:spin}, only certain computing resources will execute Line~16
based on their evaluation of the condition mentioned in Line~15. 
The operations in Line~15 represent a memory read and a memory write.  
The proportion of the memory operations and the computations performed by each resource affects and determines its delivered performance.
Consequently, the MPI rank with the largest finishing time dominates the performance of the entire process of generating the \mbox{spin-images}. 
DLS techniques are the best candidates to address the challenge of the workload imbalance.

\begin{figure}
	\centering
	\includegraphics[width=\textwidth, clip,trim=0cm 13.4cm 22cm 0cm]{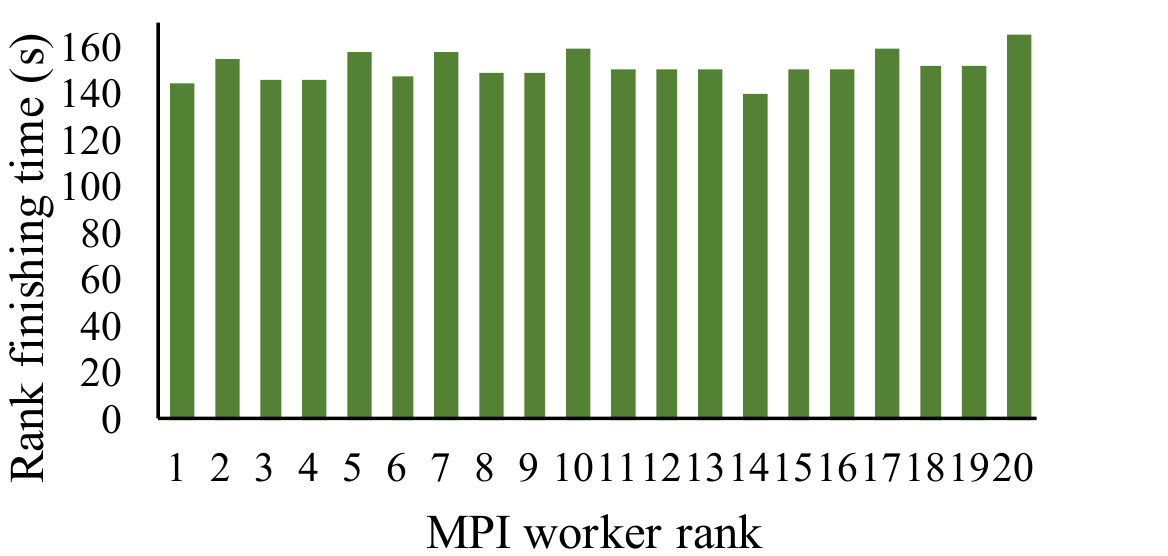}
	\caption{Variation in MPI ranks' finishing times for executing the \mbox{PSIA} application with 20 MPI worker ranks and one MPI master rank to generate 80,000 \mbox{spin-images}}
	\label{fig:loadim}
\end{figure}

\subsection{Dynamic Loop Scheduling}
\label{subsec:dls}
In scientific applications, loops are, in general, one of the main sources of parallelism. 
Parallel loops are categorized as \mbox{DOALL} and \mbox{DOACROSS} loops~\cite{chen1991empirical}. 
The \mbox{DOALL} loops have no dependencies between their iterations while \mbox{DOACROSS} loops consist of iterations that are data-dependent.
As shown in Algorithm~\ref{algo:spin}, there are no dependencies between the iterations of the outer loop~\mbox{(Lines 4-21)}.
Therefore, the \mbox{PSIA} is an example of a \mbox{DOALL} loop. 
In this section, the most common and successful dynamic loop scheduling~\mbox{(DLS)} techniques for the \mbox{DOALL} loops are discussed.
The \mbox{DLS} techniques are used to schedule loops with no dependencies between their iterations, or loops where most dependencies between iterations can be eliminated via various loop transformations.   
Using \mbox{DLS}, the scheduling decisions are performed during the application execution time~\cite{fann2000intelligent}.

The \mbox{DLS} techniques considered in this work include \mbox{SS}~\cite{tang1986processor}, \mbox{GSS}~\cite{polychronopoulos1987guided}, and \mbox{FAC}~\cite{hummel1992factoring}.
\mbox{SS} assigns a single loop iteration each time to a requesting computing resource.
The main advantage of \mbox{SS} over other \mbox{DLS} techniques is its ability to achieve an optimized load balance between all processing elements. 
However, this advantage comes at a very high overhead.
\mbox{GSS} divides the total number of loop iterations into variable-sized chunks of loop iterations.
In each scheduling step, \mbox{GSS} divides the remaining loop iterations by the total number of processing elements.
\mbox{GSS} is considered as a compromise between \mbox{SS} and \mbox{STATIC}, providing an acceptable load balance at an acceptable scheduling overhead.
\mbox{GSS} has the disadvantage of overloading  the first free and requesting computing resource with the first and largest chunk of iterations.
The remaining loop iterations may not be sufficient to ensure a balanced execution among the computing resources.
\mbox{FAC} was designed to handle iterations of variable execution time. 
It schedules the loop iterations in batches of P of equal sized chunks where P is the total number of computing resources.

The reason for selecting the three  above-mentioned \mbox{DLS} techniques and STATIC for the present work is to cover a broad spectrum of the performance of the \mbox{PSIA} using the loop scheduling techniques.
\mbox{SS} and \mbox{STATIC} represent the two extreme cases of the \mbox{DLS} techniques. 
\mbox{STATIC} has the lowest communication overhead and the lowest ability to balance the execution of the loop iterations among the workers.
SS has the highest communication overhead and the highest ability to balance the execution of the loop iterations among the workers. 
The expected performance of \mbox{GSS} and \mbox{FAC} represent intermediate points between \mbox{STATIC} and \mbox{SS}. 
Further work is needed and planned as future work to include other more complex \mbox{DLS} techniques.

\subsection{Related Work}
\label{subsec:related_work}
In~\cite{eleliemy2016loadbalancing}, an empirical approach was used to achieve the best performance of \mbox{PSIA} executing on a heterogeneous computing system that consisted of an Intel~\mbox{CPU} and an Intel Knights Corner~\mbox{(KNC)} co-processor.
The main goal of the work in~\cite{eleliemy2016loadbalancing} was to achieve a load balanced execution of the algorithm between the 24 cores CPU and the 64 cores \mbox{KNC}.
The approach taken in~\cite{eleliemy2016loadbalancing} \textit{statically} divides the workload (the generation of \mbox{spin-images}) unequally in such a way that guarantees that the \mbox{CPU} cores and the \mbox{KNC} cores finish the execution at the same time.
In practice, to perform such a static division of the generation of the \mbox{spin-images}, certain information regarding the time to generate each \mbox{spin-image} is required.
This information was obtained by generating each \mbox{spin-imag}e on the two available computing architectures.
However, the obtained information was only valid for specific computing architectures and for the input data used.
Motivated by the work in~\cite{eleliemy2016loadbalancing}, the present work demonstrates the need for using \textit{dynamic loop scheduling} within~\mbox{PSIA} and extends it into~\mbox{EPSIA}.
\mbox{EPSIA} employs dynamic loop scheduling to execute efficiently \emph{both} on heterogeneous as well as homogeneous computing resources.

For distributed memory architectures (similar to the ones used in this work), the \mbox{DLS} techniques were integrated within a \mbox{master-worker} execution model~\cite{IBto,IBtog}. 
Without loss of generality, the present work differs from~\cite{IBto, IBtog} as follows: 
(1)~The~master is a dedicated resource and performs the \mbox{DLS}-based chunk calculations and the work assignment using multiple threads (16 threads); 
(2)~There is no communication or work reassignment among the workers; 
(3)~The input data is initially replicated in the main memory of all workers;
(4)~The~workers only send the results of calculating all chunks \emph{after} they receive the termination signals from the master.
\section{The Proposed Efficient PSIA}
\label{sec:proposed}
The efficient version of PSIA, proposed in this work and denoted \mbox{EPSIA}, is introduced next. 
The \mbox{EPSIA} employs a \mbox{master-worker} execution model.
As shown in \figurename{~\ref{fig:protocol}}, the \mbox{master-worker} communication protocol consists of five steps: 
(1)~A free worker \textit{requests} an amount of work (chunk of loop iterations); 
(2)~The master \textit{calculates} (according to the selected DLS technique) and \textit{assigns} a chunk of loop iterations to the requesting worker; 
(3)~When the worker finishes the assigned chunk of loop iterations, it notifies the master and \textit{requests} another chunk of loop iterations; 
(4)~If there are still unexecuted loop iterations, the master calculates and assigns  a new chunk of loop iterations to that worker; otherwise it sends a \textit{termination} signal;
(5)~When a worker receives a \textit{termination} signal, it \textit{sends back} the results of executing the assigned chunks to the master.

\begin{figure}
	\begin{center}
		\includegraphics[width=\textwidth, clip, trim= 0cm 6.5cm 0cm 0cm]{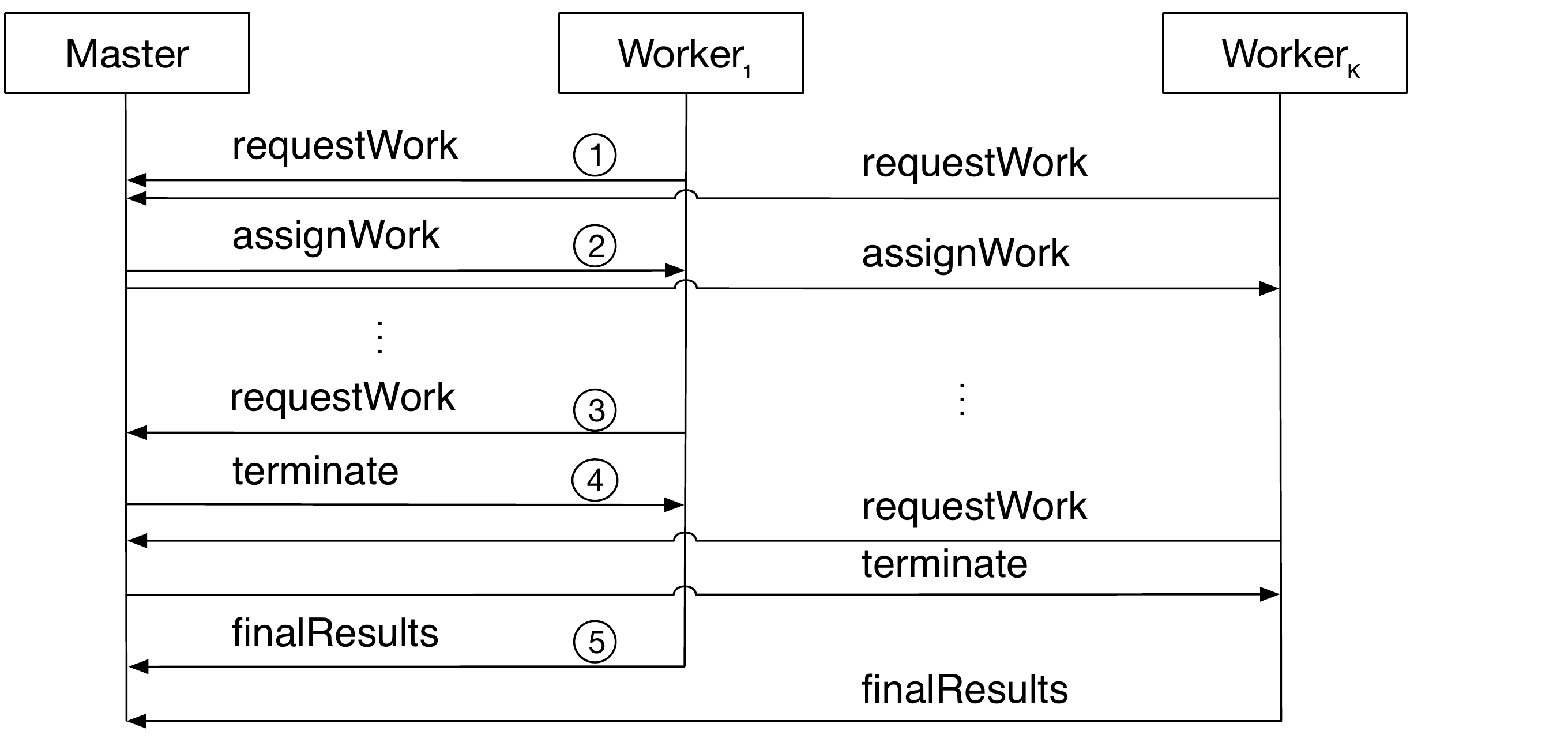}
		\caption{The master-workers communication protocol. The encircled number denotes the order of control messages exchanged }
		\label{fig:protocol}
	\end{center}
\end{figure}

To integrate the \mbox{master-worker} execution model into \mbox{PSIA}, certain changes are needed to be made to Algorithm~\ref{algo:spin}.
The proposed algorithm is shown in Algorithm~\ref{algo:adaptedcalc} in which, the code parts in blue font color~\mbox{(Lines 1, 2, and 3)} represent the modifications required for Algorithm~\ref{algo:spin} to employ the master-worker execution model.

Recall from~\ref{subsec:dls} that the current work differs from previous work~\cite{IBto,IBtog} as following:  (1)~The master is dedicated to handle the worker requests using multiple threads; (2)~The workers do not communicate with each others; (3)~The input data is replicated; (4)~The results are collected from the workers at the end. 
These distinctions are made to more closely align with the earlier \mbox{PSIA} implementation and to allow a meaningful comparison with \mbox{EPSIA}.
As discussed next in Section~\ref{sec:experimentsetup}, the main memory of recent computing resources satisfies the memory requirements of dense 3D objects. 
Therefore, replicating the information of the 3D object and storing the generated \mbox{spin-images} on the worker side result in lightweight messages between the master and the workers.
Moreover, a dedicated master resource offers rapid responses to the workers, especially when executing on large number of workers.
The usefulness of the \mbox{master-worker} execution model and the integration of the communication protocol from \figurename{~\ref{fig:protocol}} in \mbox{EPSIA} is described as two pseudo code  algorithms than can be found below.
\begin{figure}
	\removelatexerror
	\begin{algorithm}[H]
	\SetKwInOut{Input}{Inputs}
	\SetKwInOut{Output}{Output}
	adCalculateSpinImages (W, B, S, OP, M, {\color{blue}spinImages, start, end})\;
	\Input{W: \mbox{image width}, B: \mbox{bin size}, S: \mbox{support angle}, \mbox{OP: list of oriented points}, \mbox{M: number of oriented points}, \mbox{spinImages: list of spin-images to be filled}}
	 \For{{\color{blue}imageCounter = start $\rightarrow$ end}}
	{ 
		\color{black}
		{\color{blue}P = OP[imageCounter]\;}
		tempSpinImage[W, W]\;
		init(tempSpinImage)\;
		\For{j = 0 $\rightarrow$ $M$}
		{
			X = OP[j]\;
			$np_i$ = getNormal(P)\;
			$np_j$ = getNormal(X)\;
			\If{acos($np_i \cdot np_j$) $\le S$}
			{
				$k$  =  $\Bigg \lceil$ $\cfrac{W/2 - np_i \cdot (X-P) }{B}$ $\Bigg \rceil$ \;
				\vspace{0.2cm}
			   $l$ =  $\Bigg \lceil$  $\cfrac{ \sqrt{||X-P||^2 - (np_i\cdot(X-P))^2} }{B}$  $\Bigg \rceil$\;
				
				\If{0 $\le$ k $\textless$ W and 0 $\le$ l $\textless$ W}
				{ tempSpinImage[k, l]++\;	}
			}
		}
		add(spinImages, tempSpinImage)\;
	}
	\caption{Modification of the \mbox{spin-image} calculation for integration with the \mbox{master-worker} execution model and the \mbox{DLS} techniques}
	\label{algo:adaptedcalc}
\end{algorithm}
\end{figure}

\begin{algorithm}[H]
	\SetKwInOut{Input}{Inputs}
	\SetKwInOut{Output}{Output}
	generatingSpinImages (OF, W, B, S, N, DM)\;
	\Input{\mbox{OF: location of the input data}, \mbox{W: image width}, \mbox{B: bin size}, \mbox{S: support angle}, \mbox{N: number of generated spin-images}, \mbox{DM: DLS technique}}
	\Output{spinImages: list of generated spin images}
	OP = read3DPoints(OF)\;
	scheduledTasks = 0\;
	schedulingStep = 0\;
	receivedResults = 0\;
	startEnd[2]\;
	workersCount = getCountOfWorkers()\;
	sendToWorkers(OP)\; 
	\While{scheduledTasks $\textless$ N}
	{
		requestWork = receiveRequestAnyWorker()\;
		worker = getSourceOfRequest(requestWork)\;
		chunk = getChunk(DM, schedulingStep, N, workersCount)\;
		startEnd[0] = scheduledTasks\; 
		startEnd[1] = scheduledTasks + chunk\;
		sendResponse(worker, startEnd, assignWork)\;
		scheduledTasks = scheduledTasks + chunk\;
	}
	\While{receivedResults $\textless$ workersCount}
	{
		request = receiveRequestFromAnyWorker()\;
		requestType = getRequestType(request)\;
		worker = getSourceOfRequest(request)\;
		\eIf{requestType = assignWork}
		{
			sendResponseToWorker(worker, NULL, terminate)\;
		}
		{
			receiveDataFromWorker(worker, tempSpinImages)\;
			add(spinImages, tempSpinImages)\;
			receivedResults++\;
		}
	}
	\caption{The proposed \mbox{APSIA} master perspective}
	\label{algo:master}
\end{algorithm}
\newpage
\begin{algorithm}[H]
	\SetKwInOut{Input}{Inputs}
	\SetKwInOut{Output}{Output}
	generatingSpinImages (OF, W, B, S, DM)\;
	\Input{\mbox{OF: location of the input data}, \mbox{W: image width}, \mbox{B: bin size}, \mbox{S: support angle}, \mbox{DM: DLS technique}}
	\Output{\mbox{spinImages: list of generated spin images}}
	receiveFromMaster(OP)\;
	M = getLength(OP)\;
	startEnd[2]\;
	sendRequest(assignWork)\;
	response = receiveResponseFromMaster()\;
	spinImages = createSpinImagesList(M)\;
	\While{response = assignWork}
	{
		startEnd = getResponseData(response)\; 
		/* as shown in Algorithm~2 */\; 
		adCalculateSpinImages(W, B, S, OP, M, spinImages, startEnd[0], startEnd[1])\;	
	}
	sendDataToMaster(spinImages)\;

	\caption{The proposed \mbox{APSIA} worker perspective}
	\label{algo:worker}
\end{algorithm} 

\section{Setup of Experiments}
\label{sec:experimentsetup}
This section contains certain essential information concerning the experimental setup needed to reproduce the current.
\subsection{Input Data Set} 
\label{subsec:data_set}
As discussed in Section~\ref{subsec:psia}, the time complexity of \mbox{SIA} is~$O(N M)$. 
It is important to consider the \mbox{3D} objects of high density regarding the number of \mbox{3D} points.
In Table~\ref{tab:dataset}, the objects of the \mbox{3D} mesh watermarking~\cite{dataset} data set are presented. 
The \mbox{3D} mesh watermarking data set consists of ten dense \mbox{3D} objects. 
These \mbox{3D} objects vary regarding the number of points from approximately~$3K$ to approximately~$826K$ points. 

Out of the 3D~objects in Table~\ref{tab:dataset}, the \textit{Ramesses} object is considered as the extreme case in terms of \mbox{3D} points density for the \mbox{EPSIA}.
\textit{Ramesses} object contains the largest number of oriented points, approximately~$826K$, 
and is considered for comparing the performance of the proposed~\mbox{EPSIA} and the earlier~\mbox{PSIA}.
Similar to~\cite{eleliemy2016loadbalancing}, the present work considers the three \mbox{spin-image} generation parameters W, B and S to be 5, 0.1 and 2$\pi$, respectively. In addition, the present work considers the number of generated \mbox{spin-images}~N to be 10\% of the total number of oriented points~M~(see Table~\ref{tab:notation})~\cite{eleliemy2016loadbalancing}.

\begin{table}[!t]
	\caption{The 3D objects in the mesh watermarking \mbox{data set}~\cite{dataset}}
	\vspace{-0.35cm}
	\label{tab:dataset} 
	\begin{center}
		\begin{tabular}{ c | c}
			\textbf{Object} & \textbf{Approximate number of points ($\times$ $10^3$)} \\ \hline
			Cow & 3  \\ \hline
			Casting & 5  \\ \hline
			Bunny & 35  \\ \hline
			Hand & 37  \\ \hline
			Dragon & 50  \\ \hline
			Crank & 50 \\ \hline
			Rabbit & 71  \\ \hline
			Venus & 101   \\ \hline
			Horse & 113  \\ \hline
			\textbf{Ramesses} & \textbf{826}  \\ \hline
		\end{tabular}
	\end{center}
\end{table}

\subsection{Hardware Platform Specifications} 
\label{subsec:hw_specs}
Two different types of computing resources are used in this work to assess and compare the performance of the proposed \mbox{EPSIA} and the earlier \mbox{PSIA}. 
The first platform type, denoted Type1, represents a \mbox{two-socket} processor~(20~cores) Intel Xeon~\mbox{E5-2640} with~\mbox{64 GB~RAM}. Each core has 32~KB and 256~KB as L1 and L2 caches, respectively. 
Cores of the same processor socket share 25~MB L3 cache. 
The second platform type, denoted Type2, is a standalone Intel Xeon Phi~7210~(64~cores) with~\mbox{96 GB~RAM}.  Each core has 32~KB L1 cache. Each tile (two cores) has 1~MB L2 cache.
The platform types Type1 and Type2 are part of a computing cluster that consists of 26~nodes:~22 of Type1 and~4 nodes of Type2.
All nodes are interconnected in a non-blocking fat-tree topology. The network characteristics are: Intel \mbox{OmniPath} fabric, \mbox{100~GBit/s} link bandwidth, and 100~ns~(for homogeneous resources) and 300~ns~(for heterogeneous resources) link latency. 
This computing cluster is actively used for research and educational purposes.
Therefore, only eight nodes of Type1 and four nodes of Type2 were dedicated to the present work.

\subsection{Implementation and Execution Details}
\label{subsec:implementation}
The Intel message passing interface library (Intel-MPI, version \mbox{17.0.1}) was used to compile and execute the implementation of the proposed \mbox{EPSIA}.
The \mbox{Intel-MPI} library has the advantage of default pinning of operating system level processes to hardware cores (i.e., process pinning). 
Pinning a particular MPI process to a hardware core eliminates  the undesired process migration that may be performed by the operating system during execution.
Moreover, to examine the performance of the \mbox{DLS} in one of the worst cases, all master-worker control and data messages exchanged (cf. Fig.~\ref{fig:protocol}) are implemented using MPI point-to-point synchronous communication primitives.

A user-specified machine file is used to map the MPI ranks to the computing resources (cores of nodes of Type1 and Type2).
All computing resources are listed in the machine file in a certain order.
This order indicates the MPI rank assigned to each computing resource during the execution of the application.
Executing on homogeneous resources of Type1 or Type2 where all computing resources are similar, this order has no influence on performance.
However, when executing on heterogeneous resources of Type1 and Type2, all computing resources of Type2 are listed in the machine file before computing resources of Type1.
The rational behind this listing is to enable the nodes with the largest number of cores~(Type2) take the first MPI ranks.
In the next section, the influence of this listing is presented and discussed.
The master~(MPI rank~=~0) is always mapped to a dedicated computing resource. 
This computing resource is a core of a dedicated node of Type1.
This dedicated computing resource is always written at the beginning of the machine file.

Each experiment has been executed fifteen times to obtain certain descriptive measurements, such as maximum, minimum, average, median, first, and third quartiles. 

\subsection{Reproducibility Information}
\label{subsec:reprod}
To enable reproduction of this work, apart from the information in Sections~\ref{subsec:data_set},~\ref{subsec:hw_specs}, and~\ref{subsec:implementation}, the source code of the proposed \mbox{EPSIA} is available upon request from the authors under the lesser general public license~\mbox{(LGPL)}. 
In addition to the raw results which are already available online\footnote{https://c4science.ch/diffusion/3863/}, an Easybuild\footnote{http://easybuild.readthedocs.io} configuration file is provided to guarantee the usage of a toolchain that is similar to the one used for this work.
The code was compiled and executed using the Intel MPI~version~17.0.1. 
The \mbox{O3} compilation flag was used for execution on Type1 nodes. 
In addition, the \mbox{xCommon-AVX512} compilation flag was used for execution on Type2 nodes.
All parallel computing nodes use CentOS Linux release~7.2.1511 as operating system. 
The open grid scheduler/grid engine version 2011.11p1 is the batch system of the HPC platform where all experiments were performed. The network file system (NFS) version 4 (NFS4) is configured and used for the HPC platform.

\section{Experimental Results and Evaluation}
\label{sec:results}
In this section, the results of executing the \mbox{EPSIA} on homogeneous and heterogeneous computing resource are discussed and compared to the results of executing the \mbox{PSIA}.
\subsection{Performance of~\mbox{EPSIA} vs.~\mbox{PSIA} on Homogeneous Computing Resources}
\label{subsec:perfomancewod}

In this section, the performance of the \mbox{PSIA} is compared to the performance of the proposed \mbox{EPSIA} for two scalability profiles: weak and strong.
As discussed in Section~\ref{subsec:related_work}, the \mbox{PSIA} \textit{statically} divides and assigns the \mbox{spin-image} calculations to the available computing resources. 
In all experiments, the \mbox{PSIA} is referred to as \mbox{PSIA-STATIC}. 
 \mbox{EPSIA-SS}, \mbox{EPSIA-GSS} and \mbox{EPSIA-FAC} denote the proposed \mbox{EPSIA} code parallelized with the three \mbox{DLS} techniques: \mbox{SS}, \mbox{GSS}, and \mbox{FAC}, respectively.
 
\subsubsection{Weak Scalability}
\label{subsubsec:weawod}

For conducting weak scalability experiments, the number of the generated \mbox{spin-images} and the number of the computing resources are increased such that their ratio is kept constant at $8K$ \mbox{spin-images} per computing node. 
The number of the generated \mbox{spin-images} in this ratio represents approximately~1\% of the total \mbox{spin-images} that can be generated from the \textit{Ramesses} object.
This work percentage is selected to result in a suitable, yet representative, execution time per experiment, given that each experiment has been executed fifteen times. 

A comparison between the \emph{parallel execution time} of  the proposed \mbox{EPSIA} and the PSIA achieved by executing them on different node counts of the two platform types is presented in \figurename{~\ref{fig:wst1}~and~\ref{fig:wst2}}.
The execution time of the \mbox{PSIA-STATIC} is significantly higher than that of \mbox{EPSIA-SS}, \mbox{EPSIA-GSS}, and \mbox{EPSIA-FAC}. 
For \mbox{PSIA-STATIC} on \mbox{Type1} nodes, increasing the number of the generated \mbox{spin-images} from~$8K$ to~$64K$ (i.e., by a factor of 8) and increasing the number of the computing resources from~$20$ to~$160$ (i.e., by a factor of 8) result in an undesired performance degradation. 
Specifically, the execution time increased from~21 to 25~seconds, an almost~20\% increase.
The \mbox{EPSIA-SS} did not exhibit such performance degradation.
Specifically, the execution time increased from~20 to 20.5~seconds, an almost~1\% increase.
Similarly to the performance on \mbox{Type1} nodes, for \mbox{PSIA-STATIC} on \mbox{Type2} nodes, increasing the number of the generated spin images from~$8K$ to~$32K$ (i.e., by a factor of 4) and increasing the number of the computing resources from~$64$ to~$256$ (i.e., by a factor of 4) result in an undesired performance degradation. 
In particular, the execution time increased from~30 to 35~seconds, approximately a~17\% increase.
Executing the \mbox{EPSIA-SS} on \mbox{Type2} nodes resulted in poor performance compared with the execution on \mbox{Type1} nodes. 
In particular, the execution time increased from~27.5 to 30~seconds, approximately a~9\% increase.
However, \mbox{EPSIA-SS} still outperforms all other versions of the proposed \mbox{EPSIA}.

\begin{figure}
		\centering
	\includegraphics[width=\textwidth, clip,trim=0cm 2.5cm 0cm 0cm]{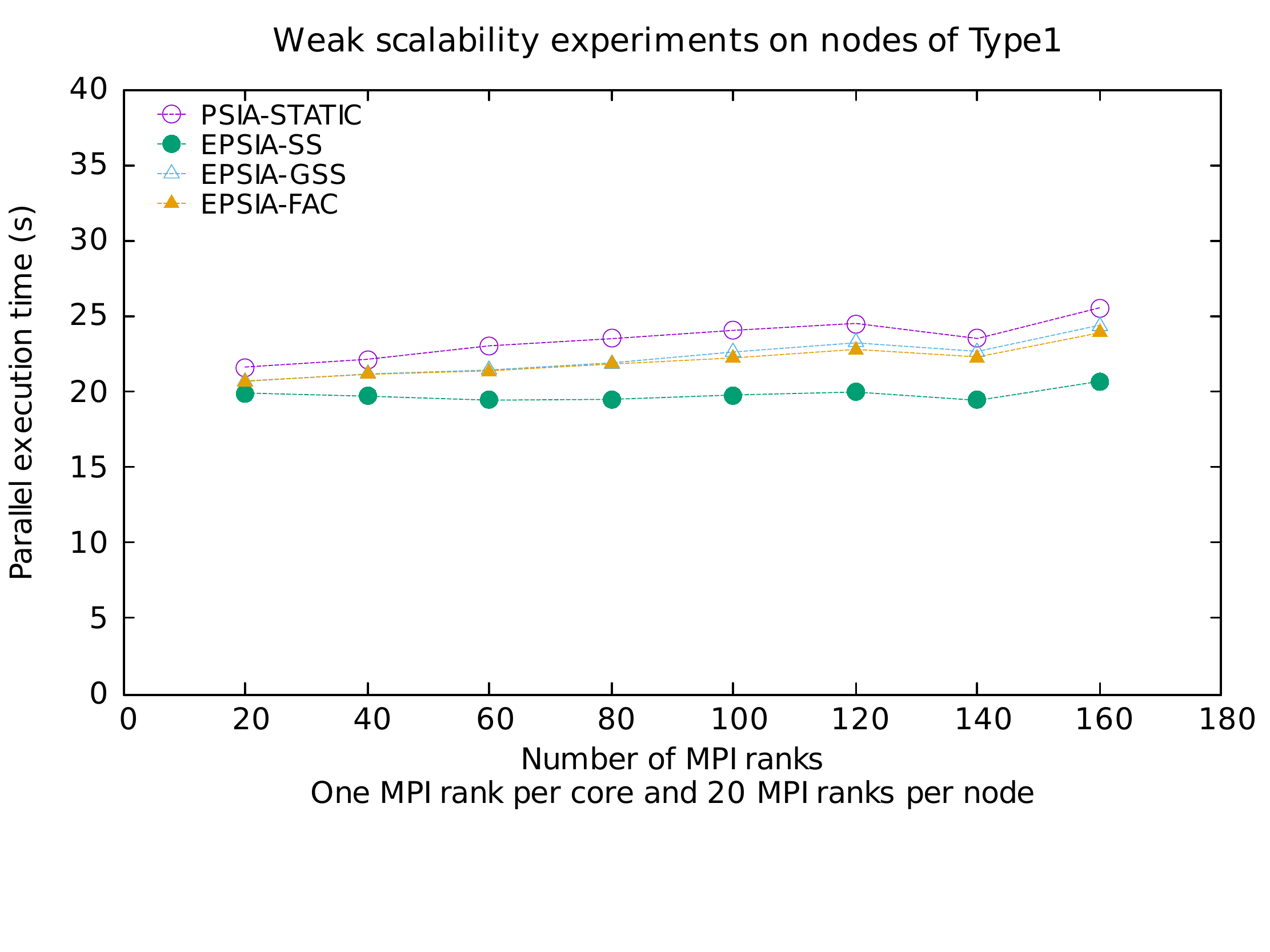}
	\caption{Scalability of the proposed \mbox{EPSIA} and the earlier \mbox{PSIA} on homogeneous computing resources of Type1. The number of generated \mbox{spin-images} per computing node is~$8K$.}
		\label{fig:wst1}
\end{figure}
	
\begin{figure}	
   \centering
	\includegraphics[width=\textwidth, clip,trim=0cm 2.5cm 0cm 0cm]{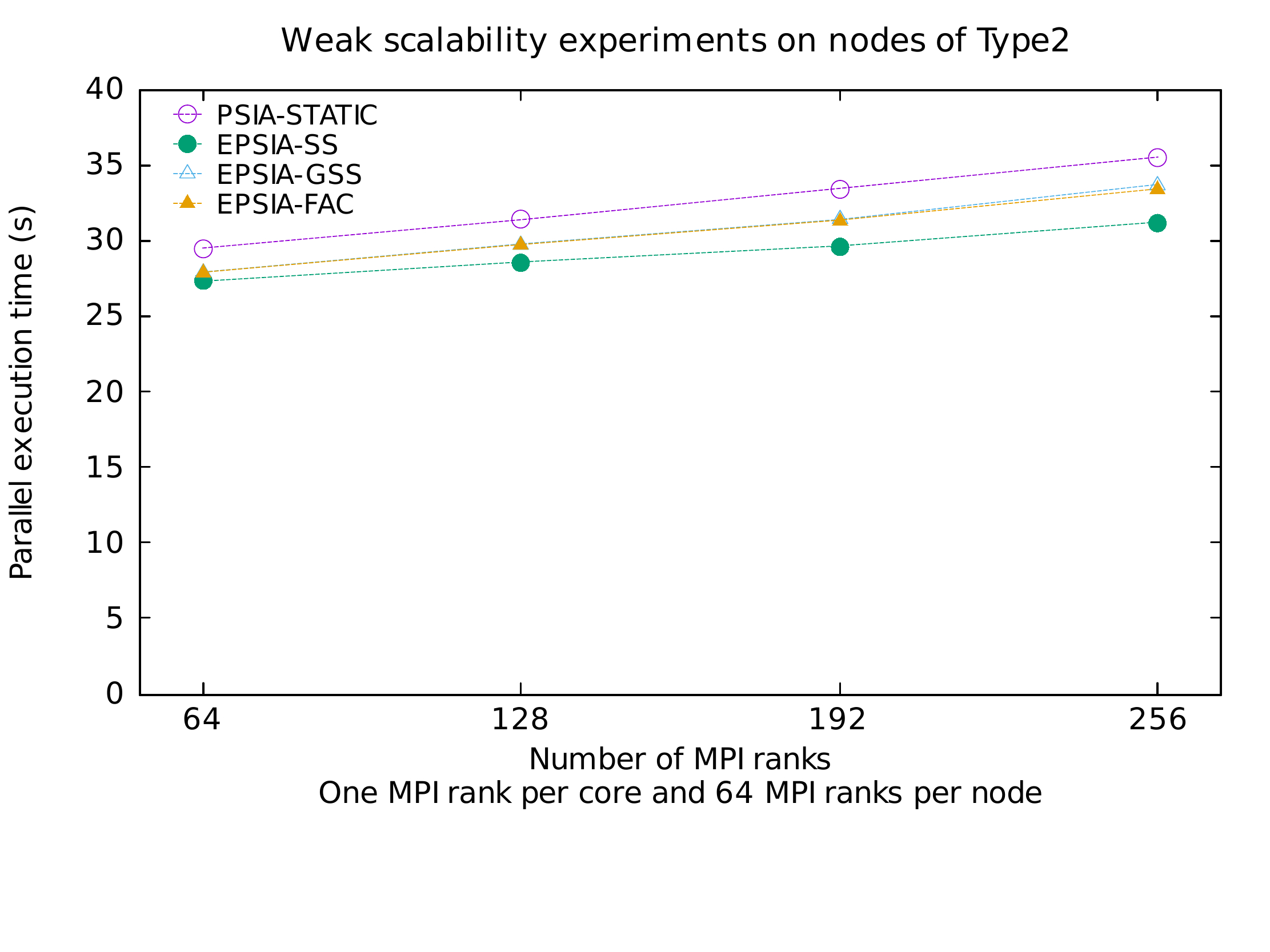}
	\caption{Scalability of the proposed \mbox{EPSIA} and the earlier \mbox{PSIA} on homogeneous computing resources of Type2. The number of generated \mbox{spin-images} per computing node is~$8K$.}
	\label{fig:wst2}
\end{figure}

Such a difference in the performance of different \mbox{EPSIA} versions and the \mbox{PSIA-STATIC} can be explained due to the load imbalance by the static division and the static assignment of the generation of the \mbox{spin-images} in  \mbox{PSIA-STATIC}.

In general, the \mbox{SS} algorithm incurs high communication overhead caused by the large volume\footnote{Depending on the input data distribution strategy, which can be either centralized, partitioned, or replicated.} and/or number of messages\footnote{At least equal to the total number of parallel tasks within the application.} between the master and the worker.
In this work, however, the input data is \emph{replicated} and the master exchanges only lightweight messages (a few bytes per message) with the workers to indicate the chunk sizes they need to execute. 
The number of such lightweight messages corresponds to the total number of chunks of tasks calculated by the dynamic loop scheduling algorithm and is different across DLS techniques. 
The superiority of the \mbox{EPSIA-SS} over the other two \mbox{EPSIA} versions can be explained by its fine-grain self-scheduled task assignment design as well as by the high speed of the network infrastructure used in the experiments and due to the usage of a multithreaded master process on a dedicated computing node. 
Both these aspects result in a more balanced execution time among the computing resources, hence, a shorter parallel execution time, using \mbox{EPSIA-SS}.

\subsubsection{Strong Scalability}
\label{subsubsec:strwod}

To perform strong scalability experiments, the number of generated spin-images is kept constant while the number of the computing resources is increased.
The number of generated \mbox{spin-images} is set at $80K$, which represents approximately~10\% of the total \mbox{spin-images} that can be generated from the \textit{Ramesses} object.

A comparison between the \emph{parallel cost} of executing the proposed \mbox{EPSIA} and the earlier \mbox{PSIA} on Type1 and Type2 nodes is presented in \figurename{~\ref{fig:st1}~and~\ref{fig:st2}}, respectively.
The parallel cost is calculated as the number of the computing resources used to execute a parallel application multiplied by the total parallel execution time of that application.
The selection of parallel cost as a performance metric (over the parallel execution time) is due to the fact that it reflects the benefits of using additional computing resources versus the time needed to execute the parallel algorithm. 
A good strong scalability profile of a program corresponds to an almost constant parallel cost for any number of computing resources. 
It can be seen in \figurename{~\ref{fig:st1}~and~\ref{fig:st2}}, that \mbox{PSIA-STATIC} does not exhibit a strong scalability profile for both computing resource types.
Similar to the weak scalability results in Section~\ref{subsubsec:weawod}, the three versions of the proposed \mbox{EPSIA}  outperform \mbox{PSIA-STATIC}. 
\begin{figure}
	\centering
	\includegraphics[width=\textwidth, clip,trim=0cm 2.5cm 0cm 0cm]{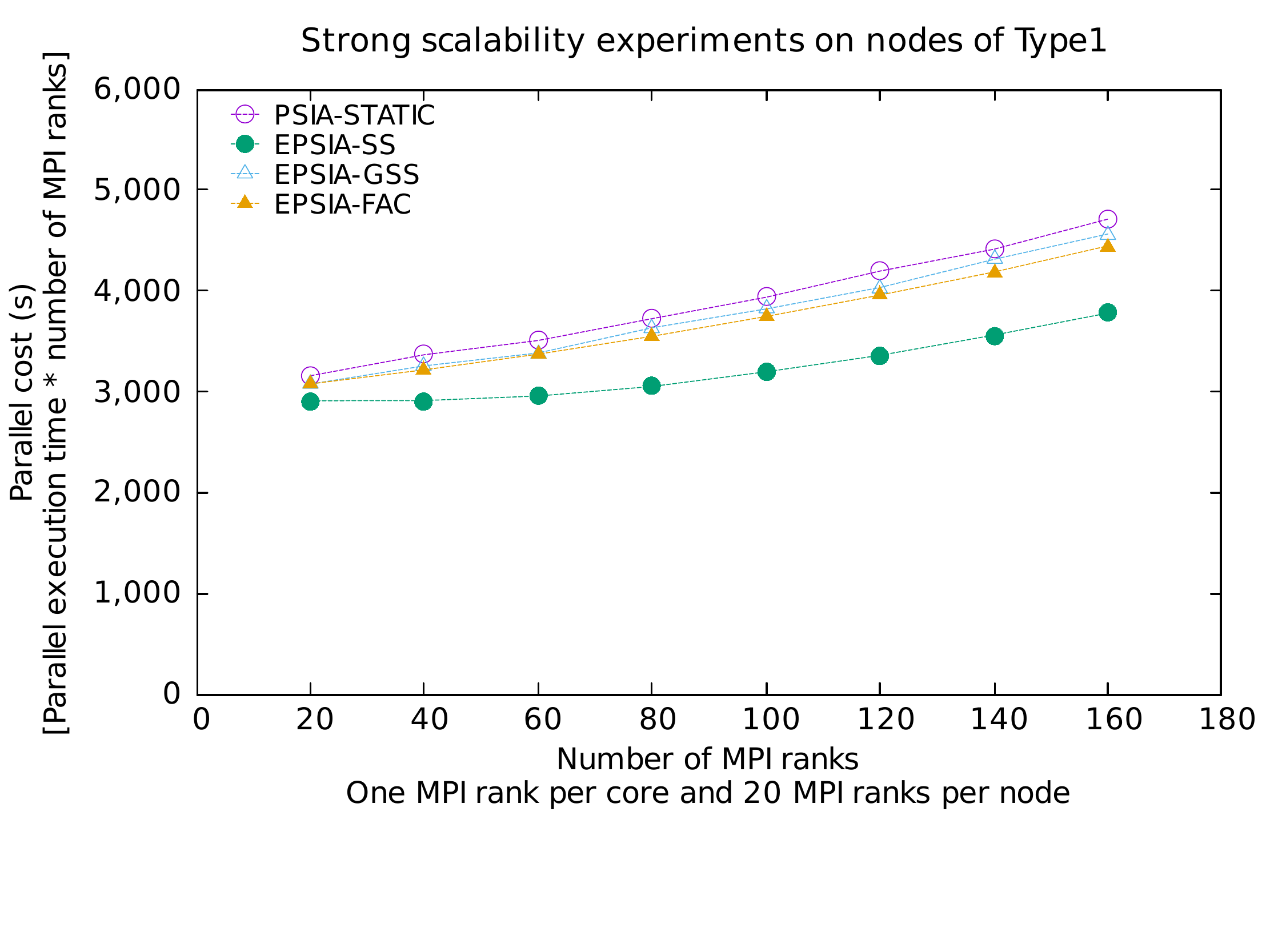}
	\caption{Scalability of the proposed \mbox{EPSIA} and the earlier \mbox{PSIA} on homogeneous computing resources of Type1. The number of generated \mbox{spin-images} is $80K$.}
	\label{fig:st1}
\end{figure}

\begin{figure}
	\centering
	\includegraphics[width=\textwidth, clip,trim=0cm 2.5cm 0cm 0cm]{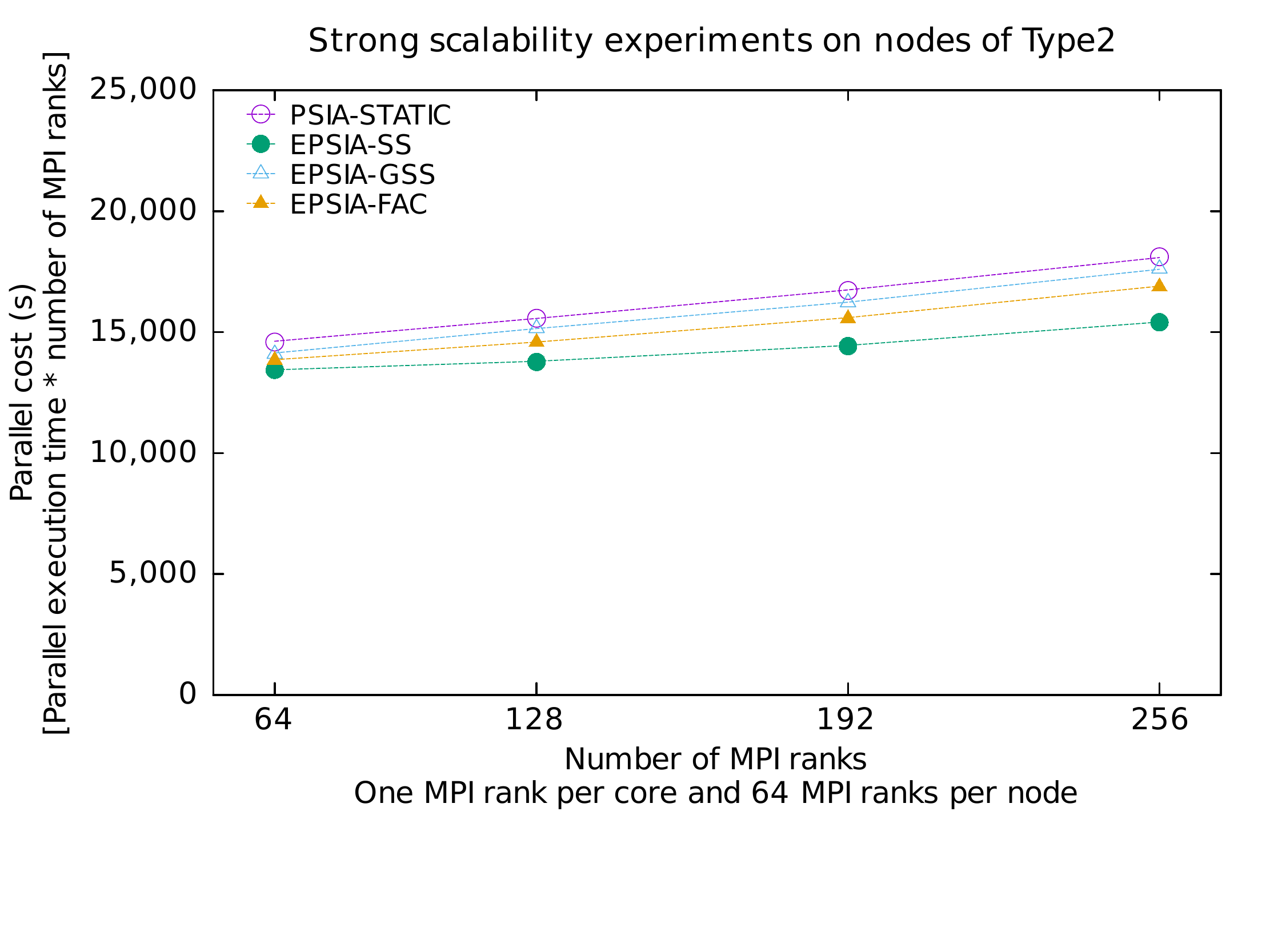}
	\caption{Scalability of the proposed \mbox{EPSIA} and the earlier \mbox{PSIA} on homogeneous computing resources of Type2. The number of generated \mbox{spin-images} is $80K$.}
	\label{fig:st2}
\end{figure}

The performance advantage of \mbox{EPSIA-SS} over \mbox{EPSIA-GSS} and \mbox{EPSIA-FAC} is attributed to the small message sizes exchanged between the master and the workers and to the high speed of the network infrastructure used in the experiments. 
The performance gap between the \mbox{EPSIA-SS}, and \mbox{EPSIA-GSS} and \mbox{EPSIA-FAC} can be explained similarly to the performance gap between the same algorithms in the weak scalability experiments in Section~\ref{subsubsec:weawod}.
The performance gap may, however, be reduced in certain other cases where the network infrastructure has a lower performance than the one used in this work. 
In both weak (Section~\ref{subsubsec:weawod}) and strong  (Section~\ref{subsubsec:strwod}) scalability experiments, the \mbox{EPSIA-SS} achieves a speed up of approximately 1.26 on the largest number of computing resources, compared to the performance of \mbox{PSIA-STATIC} on \mbox{Type1} nodes. 
On \mbox{Type2} nodes, \mbox{EPSIA-SS} achieves a speed up of approximately 1.16 compared against \mbox{PSIA-STATIC}.

\subsection{Performance of~\mbox{EPSIA} vs.~\mbox{PSIA} on Heterogeneous Computing Resources}
\label{subsec:perfomancewd}

The performance of the weak scalability and the strong scalability experiments executed on heterogeneous computing resources is shown in \figurename{~\ref{fig:hybridw}~and~\ref{fig:hybridws}}, respectively. 
These performance results are very similar to the results obtained on homogeneous computing resources. 

\mbox{EPSIA-GSS} exhibits an interesting behavior on heterogeneous computing resources compared to that on homogeneous computing resources. 
In particular, its performance is almost similar to the performance of \mbox{PSIA-STATIC}. 
This is due to the order in which the available Type1 and Type2 resources request work from the master. 
As discussed in Section~\ref{sec:background}, the \mbox{GSS} algorithm assigns the largest chunk of loop iterations to the first requesting worker. 
Recall from Section~\ref{subsec:implementation} that the heterogeneous worker computing resources listed in the machine file used in this work commence with Type2 followed by Type1. 
Also, the master is a dedicated computing resource (core) mapped on a separate node of Type1 and it is always written in the machine file before the worker computing resources of Type1 and Type2.
This listing of resources in the machine file is meant to enable the use of the computing nodes with the largest number of computing cores, i.e., 64~cores for Type2 compared to 20~cores for Type1.
Changing this listing may enhance the performance of \mbox{EPSIA-GSS} without changing the main semantic and trend of the results where \mbox{PSIA-STATIC} and \mbox{EPSIA-SS} perform the worst and the best, respectively.

In both the weak and the strong scalability experiments, the \mbox{EPSIA-SS} achieves a speed up of approximately 2 on the largest number of computing resources, compared to the \mbox{EPSIA-STATIC} on nodes of \mbox{Type1} and \mbox{Type2}.

\begin{figure}
	\centering
	\includegraphics[width=\textwidth, clip,trim=0cm 1.5cm 0cm 0cm]{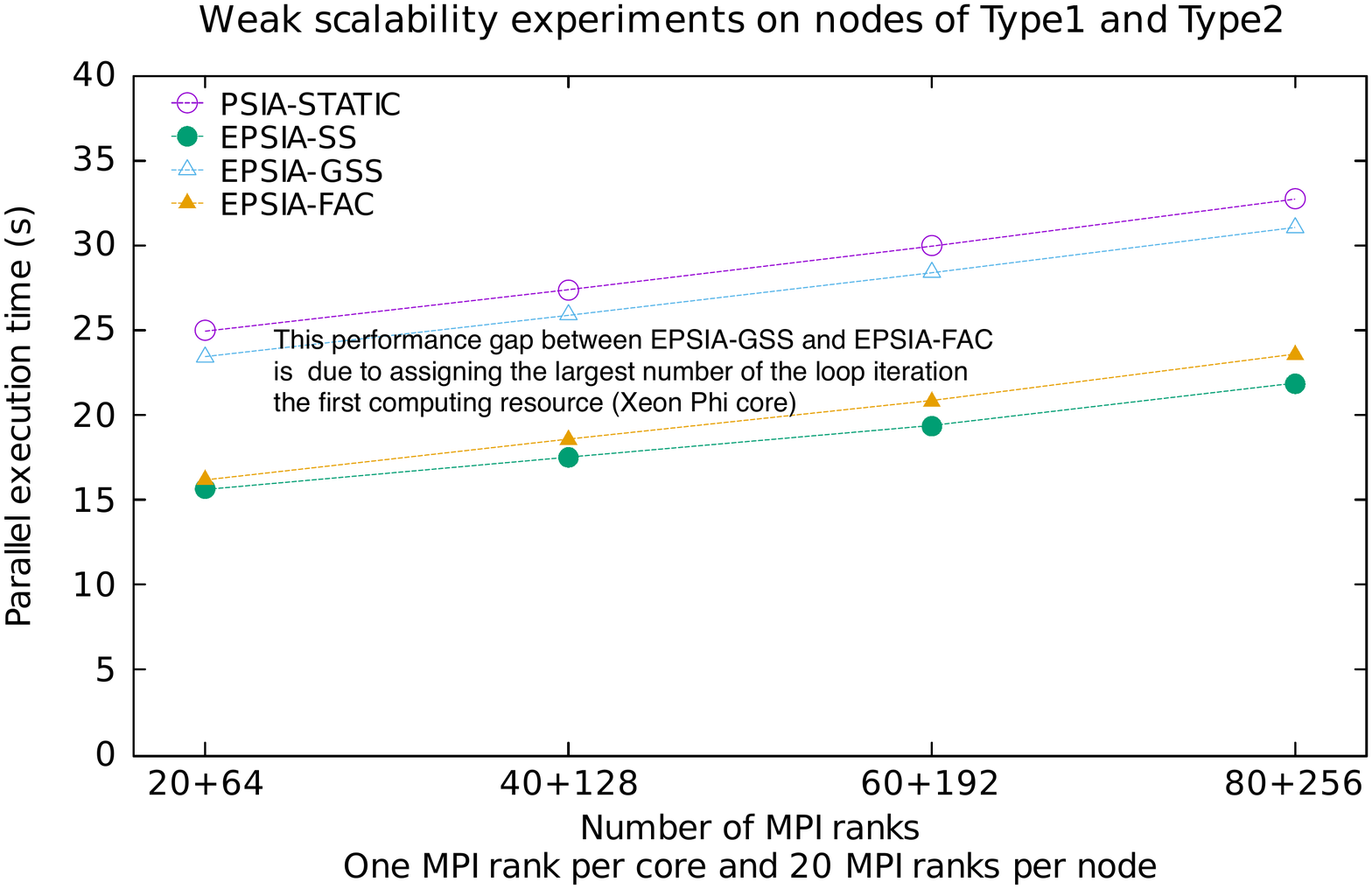}
	\caption{Scalability of the proposed \mbox{EPSIA} and the earlier \mbox{PSIA} on heterogeneous computing resources of Type1 and Type2, respectively. The number of generated \mbox{spin-images} per computing node is~$8K$.}
	\label{fig:hybridw}
\end{figure}

\begin{figure}
	\centering
	\includegraphics[width=\textwidth, clip,trim=0cm 1.5cm 0cm 0cm]{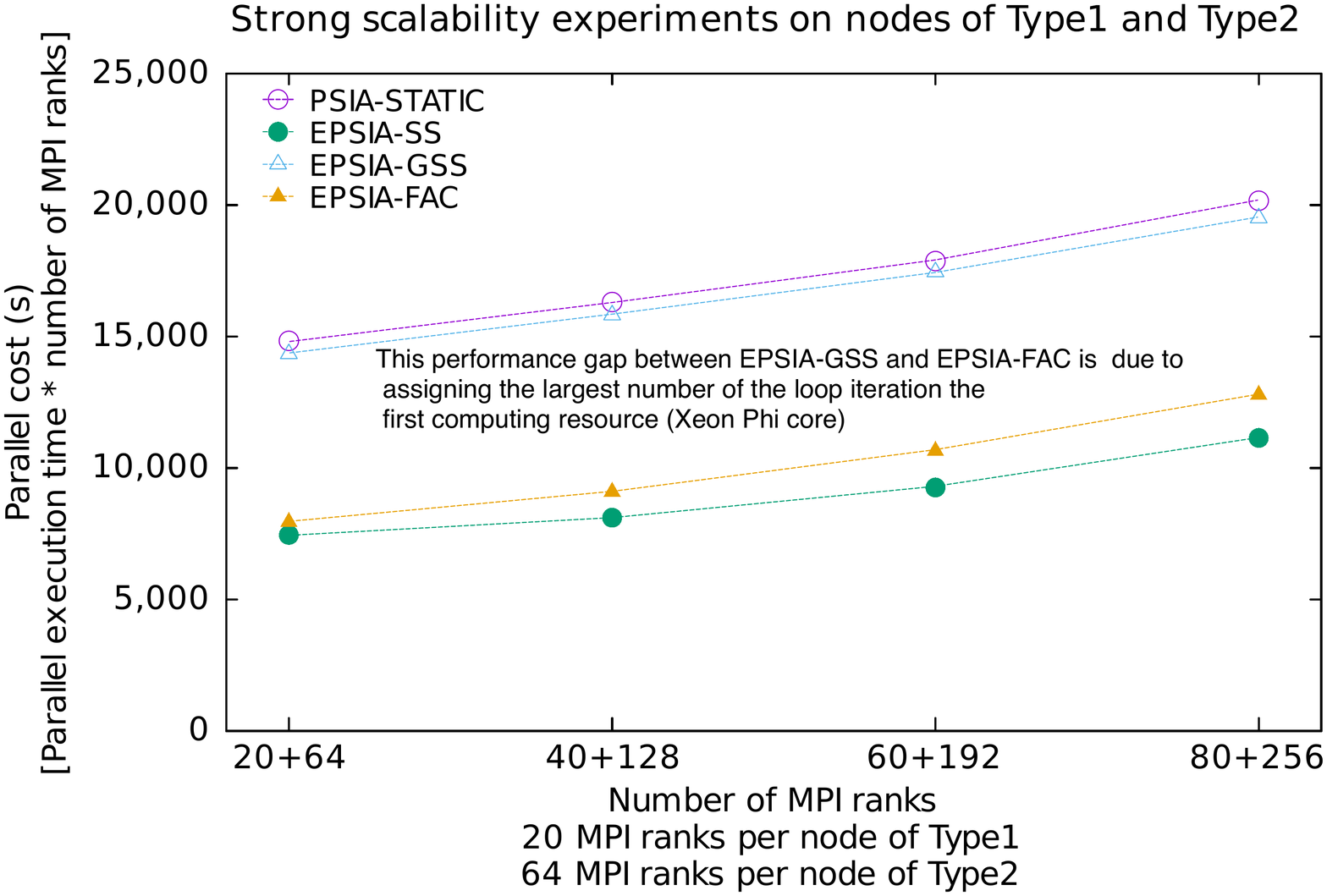}
	\caption{Scalability of the proposed \mbox{EPSIA} and the earlier \mbox{PSIA} on heterogeneous computing resources of Type1 and Type2, respectively. The number of generated \mbox{spin-images} is $80K$.}
	\label{fig:hybridws}
\end{figure} 	

\section{Conclusion and Future Work}
\label{sec:conclusion}
The static assignment of the spin-image generation tasks using \mbox{PSIA}~\cite{eleliemy2016loadbalancing} causes severe load imbalance during execution.
The load imbalance worsens when executing the \mbox{PSIA} on heterogeneous computing resources.
By employing dynamic loop scheduling (DLS) techniques and the master-worker execution model, the proposed \mbox{EPSIA} reduces the load imbalance when executing on homogeneous and on heterogeneous computing resources, as well as delivers a high performance at increased scales than in the previous work.
The proposed \mbox{EPSIA} employs three different \mbox{DLS} techniques: \mbox{SS}, \mbox{GSS}, and \mbox{FAC}. 
For the largest problem size ($80K$~\mbox{spin-images}), the performance of the \mbox{EPSIA-SS} outperforms the performance of the earlier \mbox{PSIA} by a factor of 1.2~and 2~on homogeneous and heterogeneous computing, respectively. 
Due to the high speed network used in this work, the \mbox{EPSIA-SS} shows the best performance. 
More investigation is needed and planned to assess the performance of the proposed \mbox{EPSIA} across different hardware setups.
Also, additional and more complex \mbox{DLS} techniques will be integrated with the \mbox{EPSIA}. 
As discussed in Section~\ref{subsec:perfomancewd}, the performance of the \mbox{EPSIA-GSS} is affected on heterogeneous computing resources by the type of resource requesting work in the initial chunk allocations.  
Further work is, therefore, needed to understand the effects of different resource listings in the machine file on the performance of \mbox{EPSIA}.

\section*{Acknowledgment}
This work is in part supported by the Swiss National Science Foundation in the context of the “Multi-level Scheduling in Large Scale High Performance Computers” (MLS) grant, number 169123. 
\bibliographystyle{ieeetr}
\bibliography{psia}

\end{document}